\documentclass[preprint,journal]{vgtc}       

\ifpdf
  \pdfoutput=1\relax                   
  \usepackage{graphicx}                
  \DeclareGraphicsExtensions{.pdf,.png,.jpg,.jpeg} 
\else
  \usepackage{graphicx}                
  \DeclareGraphicsExtensions{.eps}     
\fi%

\graphicspath{{figures/}{pictures/}{images/}{./}} 

\PassOptionsToPackage{warn}{textcomp}  
\usepackage{textcomp}                  
\usepackage{times}                     
\usepackage{cite}                      
\usepackage{tabu}                      
\usepackage{booktabs}                  
\usepackage{xspace}

\usepackage{paralist}
\usepackage{enumitem}
\usepackage{gensymb}
\usepackage{amsmath}
\usepackage{amssymb}
\usepackage{tikz}
\usepackage{algorithm}
\usepackage{algpseudocode}
\usepackage[version=4]{mhchem}

\usepackage{moresize}

\usepackage{wrapfig}
\usepackage{subcaption}

\DeclareMathOperator*{\argmin}{arg\,min}

\usepackage{dsfont} 
\usepackage{hyperref} 

\usepackage{listings}
\newlength{\listingindent}                
\lstdefinelanguage{goh}
{morekeywords={in,&,<,>,=},
sensitive=false,
morecomment=[l]{//},
morecomment=[s]{/*}{*/},
morestring=[b]",
}

\onlineid{1568}

\vgtccategory{Research}
\vgtcpapertype{system}

\definecolor{mred}{rgb}{.80,.12,.30}
\definecolor{MRED}{rgb}{.80,.12,.30}
\definecolor{grey}{rgb}{0.5,0.5,0.5}
\definecolor{lgrey}{rgb}{0.7,0.7,0.7}
\definecolor{purple}{rgb}{.75,0,.85}
\definecolor{pistachio}{rgb}{0.58, 0.77, 0.45}
\definecolor{myorange}{rgb}{0.94, 0.36, 0.13}
\definecolor{mygreen}{rgb}{0.08, 0.54, 0.11}

\newif\ifnotes
\notestrue

\let\origcite\cite
\renewcommand{\cite}[1]{\ifnotes\mbox{\origcite{#1}}\else \origcite{#1}\fi}

\usepackage[activate={true,nocompatibility},final,tracking=true,kerning=true,spacing=true,factor=900,stretch=25,shrink=10]{microtype}

\newcommand{\sys}{{DimBridge}\xspace}
\newcommand{\alg}{{Predicate Regression}\xspace}

\title{\sys: Interactive Explanation of Visual Patterns in Dimensionality Reductions with Predicate Logic}

\author{Brian Montambault, Gabriel Appleby, Jen Rogers, Camelia D. Brumar, Mingwei Li, Remco Chang}
\authorfooter{
\item
Brian Montambault is with Tufts University. E-mail: brian.montambault@tufts.edu.
}

\shortauthortitle{Montambault \MakeLowercase{\textit{et al.}}: \sys}

\abstract{
Dimensionality reduction techniques are widely used for visualizing high-dimensional data.
However, support for interpreting patterns of dimension reduction results in the context of the original data space is often insufficient.
Consequently, users may struggle to extract insights from the projections.
In this paper, we introduce \sys, a visual analytics tool that allows users to interact with visual patterns in a projection and retrieve corresponding data patterns.
\sys supports several interactions, allowing users to perform various analyses, from contrasting multiple clusters to explaining complex latent structures.
Leveraging first-order predicate logic, \sys identifies subspaces in the original dimensions relevant to a queried pattern and provides an interface for users to visualize and interact with them.
We demonstrate how \sys can help users overcome the challenges associated with interpreting visual patterns in projections.
}


\CCScatlist{ 
 \CCScat{K.6.1}{Management of Computing and Information Systems}%
{Project and People Management}{Life Cycle};
 \CCScat{K.7.m}{The Computing Profession}{Miscellaneous}{Ethics}
}

\teaser{
 \centering
 \includegraphics[width=\linewidth]{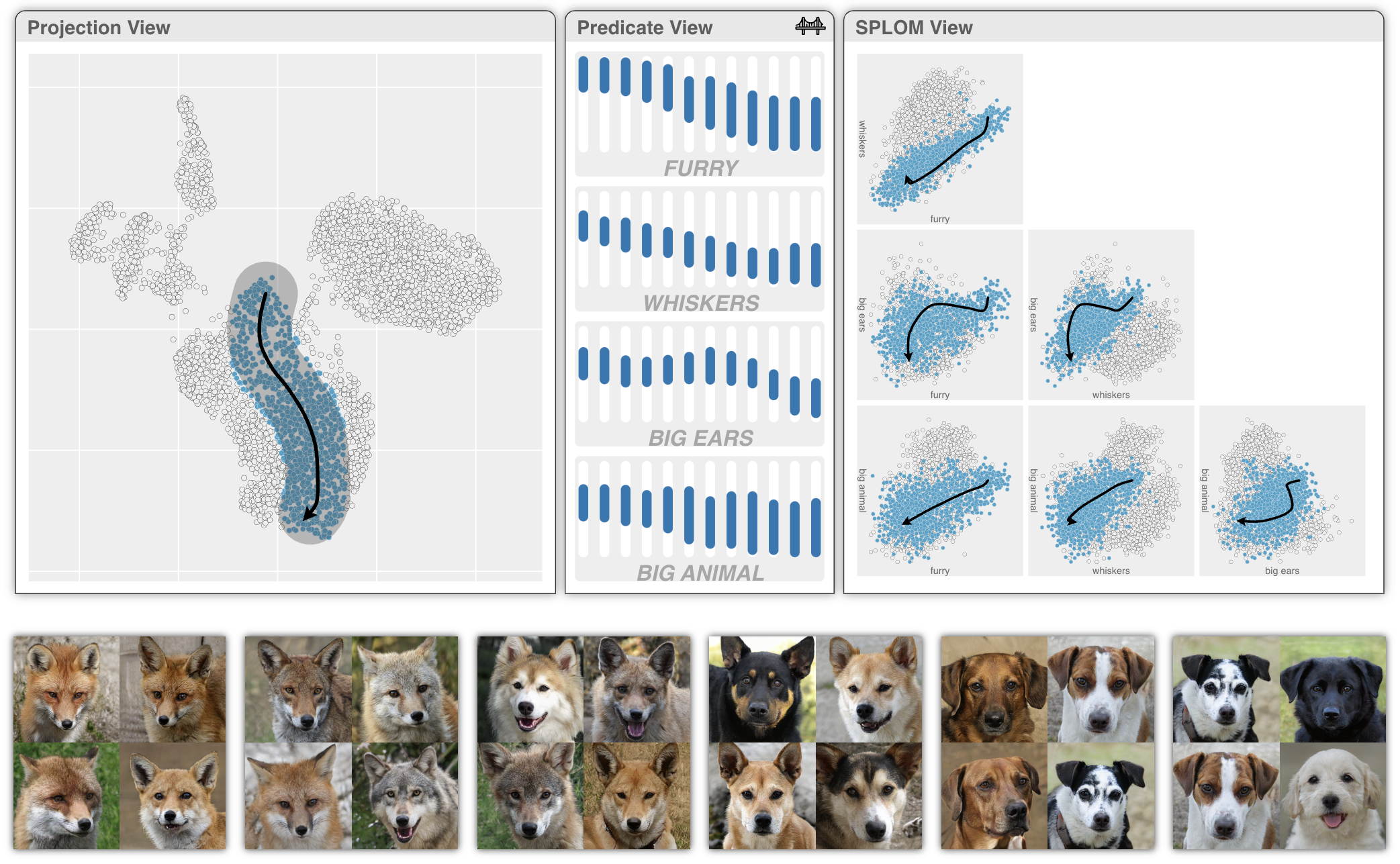}
 \vspace{-10pt}
 \caption{\sys is a system that helps users understand visual patterns in dimensionality reduction-based 2D projections. Within an interface, users can brush perceived patterns in a projection (left), and \sys computes \emph{predicates} in response (middle) -- compact explanations of the user's selection expressed in terms of the data space. The data dimensions that compose a predicate help place the data in a context well-understood by users, shown as a scatterplot matrix (right). 
 }
	\label{fig:teaser}
}

\vgtcinsertpkg

\begin{document}

\abovedisplayskip=6pt
\abovedisplayshortskip=0pt
\belowdisplayskip=6pt
\belowdisplayshortskip=6pt


\firstsection{Introduction}
\maketitle 

The demand for interactive visual exploration techniques for high-dimensional data has grown significantly, given the substantial increase in the generation of such data over the last decade\cite{liu:2017:viz_high_d_survey}. 
Among these techniques, dimensionality reduction (DR) stands out as one of the most prominent methods for visualizing high-dimensional data. 
Well-known DR techniques such as MDS\cite{kruskal1978multidimensional}, PCA\cite{pearson:1901:pca}, t-SNE\cite{maaten:2008:tsne}, and UMAP\cite{mcinnes:2018:umap} are capable of generating low-dimensional projections of data, where the spatial proximity of points encodes a measure of similarity between data items\cite{nonato:2018:projection_for_visual_analytics,espadoto:2019:quantiative_dr_survey}. 
These techniques offer 2D representations of high-dimensional data, facilitating users in identifying data patterns in the high-dimensional space more readily.

However, despite the utility of DR methods in understanding \emph{what} data items are similar, few provide insight into \emph{why} the data are similar.
As a result, users often need to rely on patterns that they perceive in the projection, e.g., clusters of points, spatial density, or distinctive shapes, to infer characteristics of the original high-dimensional data\cite{sarikaya2017scatterplots, brehmer:2014:dim_reduction_tasks, sedlmair13}.
While some of these patterns may indeed reflect underlying data characteristics, they can also be influenced by noise, artifacts stemming from the DR process\cite{martins14, nonato:2018:projection_for_visual_analytics, kobak2019art}, or even the propensity of humans to see visual patterns where none exist\cite{wattenberg:2016:using-tsne-effectively, ellis2018cognitive, wagemans2012century}.
Modern visualization tools and interaction techniques offer some assistance in understanding projections, such as helping users discern DR distortions\cite{aupetit:2007:visualizing_distortions_in_projections, lespinats:2011:checkviz, stahnke:2015:probing_projections}, examining or contrasting perceived clusters\cite{liao:2018:cluster_based_visual_abstraction, marcilio:2021:contrastive_analysis_scatterplot_dr, eckelt:2023:relationships_and_structure_embeddings}, or generating new plausible data points in the projection\cite{amorim:2012:ilamp, amorim:2015:rbf, espadoto:2023:unprojection}.


In this paper, we present \sys, building upon this existing work to contribute a new way to explore and understand perceived visual patterns in a DR projection.
The workflow for \sys is designed to be as simple as possible.
As shown in Figure~\ref{fig:teaser}, a user interactively highlights a pattern in the projection space and \sys (1) identifies a subset of data dimensions and (2) an interval of values for each dimension that concisely explain the user-selected visual pattern.
The resulting subspace, defined by the selected data dimensions and intervals, is then visualized using a scatterplot matrix (SPLOM).
This visualization enables a user to observe the selected visual pattern in a reduced context of the original high-dimensional data space -- a version of the task known as \textit{``Mapping Synthesized [Data] to Original Dimensions''} in DR task taxonomies\cite{brehmer:2014:dim_reduction_tasks, nonato:2018:projection_for_visual_analytics}.
By connecting the observed visual patterns in the projection space to the original data space, \sys helps users understand the meaning of a visual pattern in a familiar context.


\sys supports several user interactions (select, select and contrast, select and draw) for identifying visual patterns in a projection, including clusters, outliers, and shapes.
Once a pattern is selected, \sys uses specially designed predicate induction algorithms to learn the subset of data dimensions and an interval for each dimension that most effectively explains the visual pattern.
\sys is equipped with two predicate induction algorithms, a novel \alg algorithm tailored for speed and scalability, complemented by the PIXAL algorithm for accuracy\cite{montambault:2022:pixal}. Both algorithms allow for a smoothness constraint, which enables \sys to support a novel select and draw interaction.
Within this framework, predicates serve as a conceptual bridge connecting the low-dimensional (i.e., the projection space) and the high-dimensional (i.e., the original data space) space.

To demonstrate \sys's efficacy, we showcase the system on numerous datasets, demonstrate its utility within a case study, and iteratively evaluate its functionality with domain experts. 
With these examples, we also hope to demonstrate that \sys provides a means for users to explain projections using the original data space, solving key challenges in interpretability associated with conventional projection-based high-dimensional data visualization.

In summary, our work makes the following contributions:
\begin{itemize}
    \item We introduce \sys, a system that uses predicate logic to \textit{bridge} the projection and the data spaces, helping users to make sense of patterns in 2D projections of high-dimensional data.
    \item We introduce a new interaction design for selecting visual patterns within DR projections and a new algorithm for generating predicates with smoothness constraints given a selected visual pattern (\alg). 
    \item We evaluate \sys through several showcases with different domain applications, a case study, and evaluations with researchers from materials science and pharmaceutical drug discovery.
\end{itemize}


\section{Related Work}

Our work on DimBridge is based on prior work on (1) high-dimensional data visualization, (2) visualization systems that help users make sense of projections, and (3) the use of predicates in the visualization community.
\subsection{High-Dimensional Data Visualization}
\label{subsec:related-highd}
Various techniques for multi-dimensional visualization have been presented over the years. A more thorough discussion of these techniques can be found in the following surveys\cite{hoffman:2002:survey_of_vis_for_high_d, oliveira:2003:viz_high_d_survey, liu:2017:viz_high_d_survey}. We investigate these approaches to multi-dimensional visualization from the perspective of their level of dimensional reduction, from no reduction to significant reduction, and the techniques' respective affordances.

One of the most common families of methods is multiline graphs\cite{andrews:1972:andrews_curve, inselberg:1985:parallel_coords}, which plot several to many features either overlaid or stacked vertically over another dimension. This visual technique does not reduce the dimensionality of the data in terms of losing information or combining dimensions.
An example of this is the parallel coordinates plot\cite{inselberg:1985:parallel_coords,heinrich2013state}, which puts each dimension on a separate axis and draws a line connecting these axes for each instance. 
Parallel coordinates are often used in concert with sophisticated interaction techniques for selecting groups of data items, e.g., angular brushing of dimensions\cite{hauser2002angular}, multi-way brushing for high dimensional pattern discovery\cite{roberts2018smart}, as well as augmented designs that better convey relationships between dimensions\cite{bok2020augmenting}.
The multiline approach allows for the visualization of all dimensions but can become cluttered and hard to interpret as the number of dimensions increases.

In the same ethos as multiline approaches, small multiples techniques such as the scatterplot matrix (SPLOM)\cite{hartigan:1975:splom}, permutation matrices of bar charts\cite{bertin:1983:semiology}, and histogram matrices, are techniques used to display and analyze the relationships between multiple variables in a dataset.
Small multiples are closely related to the idea of dimension stacking\cite{leblanc:1990:dim_stacking}, wherein each dimension is broken into histogram-like buckets, with further dimensions recursively partitioned within preceding dimension buckets.
Small multiple views allow for a direct comparison across different data dimensions, enhancing our ability to discern patterns and anomalies within the actual data space.
Faceting data, or dimensions, across multiple views quickly leads to scalability issues, and thus methods for selecting informative views\cite{wilkinson2005graph}, used in conjunction with visual quality metrics to rank views\cite{bertini2011quality}, are common approaches to handle such problems.
This is especially the case in high-dimensional data.
However, determining the most relevant subset remains a significant challenge.
\sys uses first-order predicates to display the most relevant subset of dimensions, getting the most out of high-dimensional visualizations, which we discuss in further detail in the subsequent section \ref{sec:engine}. 


Projections, also known as dimensionality reduction (DR) methods, scale well in terms of the number of samples and dimensions, allowing them to overcome the limitations of many other high-dimensional visualization techniques.
Dimensionality reduction techniques attempt to maintain the underlying structure of the original dataset in a lower-dimensional projection. 
PCA\cite{pearson:1901:pca}, t-SNE\cite{maaten:2008:tsne}, and UMAP\cite{mcinnes:2018:umap} are three of the most popular dimensionality reduction techniques.
t-SNE and UMAP are unique from PCA in that they are non-linear DR techniques\cite{maaten:2008:tsne,mcinnes:2018:umap}, and beneficial when the data is too complex to adequately project with linear methods. 
Dimensionality reduction methods allow for detected visual patterns in a lower-dimensional space, facilitating the discovery of high-dimensional patterns.
While these methods excel at detecting patterns, making these patterns interpretable to humans is remains a challenge.

\subsection{Making Sense of Projections}
\label{subsec:related-sense}

Non-linear DR, increasingly incorporated into visual analytic systems\cite{sacha:2016:dim_reduction_interaction_survey}, aims to translate high-dimensional similarities, such as the proximity of data items to their nearest neighbors, into spatial closeness within the 2D projection space.
Spatial proximity alone, however, lacks the context for \emph{why} points are positioned, e.g., what makes 2 data items similar?
Methods exist for bringing in the context of individual data dimensions, e.g., per-dimension aggregations arranged in multiple views\cite{stahnke:2015:probing_projections, eckelt:2023:relationships_and_structure_embeddings}, dimension-specific spatial groupings in the projection\cite{sohns2021attribute}, or augmenting the projection space with 2D lines\cite{cavallo:2018:praxis} or curves\cite{faust:2018:dim_reader} that encode large variations in the data space.
Another line of work supports users in making better sense of DR hyperparameters\cite{appleby2022hypernp} rather than patterns in projections or inverting the projection to increase understanding\cite{amorim:2012:ilamp, amorim:2015:rbf, espadoto:2023:unprojection}.
There has also been an effort to add features into the DR\cite{li:2023:incorporation_human_into_embedding, fujiwara:2023:feature_learning_for_dr}.
There is also work that tends to focus on educating users about projections and hyperparameters\cite{cutura:2018:viscoder} or combining many techniques to help users find a good projection\cite{chatzimparmpas:2020:tvisne, seo:2005:rank-by-feature}.

The motivation for better understanding a projection is often driven by the visual patterns found within the scatterplot, such as the meaning behind separate clusters
\cite{wenskovitch2017towards}.
This objective is achieved by explaining clusters or classes through feature importance\cite{pagliosa:2016:understaning_attribute_variability, marcilio:2021:explaining_dr_shapley, joia:2015:representative_groups_in_projections, rauber:2015:image_feature_selection_using_dr} which can help users find similar or dissimilar groups\cite{zegarra:2020:visual_exploration_of_ratings}.
Additionally, there are numerous methods using a projection as a scaffold for interactively selecting clusters and in turn visually comparing clusters\cite{cavallo2018clustrophile} or performing a contrastive cluster analysis\cite{fujiwara:2020:dr_results_contrastive,fujiwara:2022:interactive_dr_for_comparative}.
However, as DR methods are prone to error, cluster analysis methods that convey the distortion induced by DR\cite{sun2021evosets,jeon2022distortion} remain important for proper interpretation.
Observations made through cluster analysis inform methods for steering DR\cite{Xia:2023:InteractiveContrastiveDR}, often performed in response to imperfections found in projections, reflecting an iterative process of (1) visual analysis of projections and clusters and (2) data annotation.

Many previous techniques rely on brushing to help users understand a projection.
These techniques vary from traditional brushing over the DR scatterplot\cite{fisherkeller:1975:prim9} to methods that allow users to explore the space through SPLOMs\cite{cleveland:1987:brushing_scatterplots, becker:1987:dynamic_graphs} or PCPs\cite{inselberg:1985:parallel_coords, roberts:2019:smart_brush_pcp, hauser:2002:angular_brushing_pcp}.
There are also more advanced methods such as\cite{aupetit:2014:multidim_brush_scatter, martin:1995:high_d_brushing_multivariate_data}  that allow for multiple dimensions to be brushed at once.
More recently, Jeon et al.\cite{jeon2022distortion} introduced distortion-aware brushing that resolves distortions of points local to brushed area. 

\sys builds upon these earlier methods, adding a new interaction to explain arbitrary shapes, and placing predicates front-and-center as the bridge between the dimensionality reduction result and the original data space.
\sys illuminates distortions through the coloring produced as a by-product of the predicates and uses brushing to allow the user to ask the system to explain a shape or trail within the projection.

\subsection{Predicates in Visual Analytics}
\label{subsec:related-predicates}

Predicates are frequently used in the visual analytics community to identify important subsets of data in an interpretable way, with a variety of approaches. 

Examples such as the SEER or Scorpion systems leverage predicate logic to describe user-specified patterns\cite{hanafi2017seer, garg2008model,xiao2006enhancing}, outliers\cite{wu2013Scorpion}, or query results\cite{wang:2023:DPXPlain}. 
Systems such as SuRE and Rule Matrix leverage rule-based logic for interpreting complex machine learning results\cite{yuan:2024:SuRE, Ming2019RuleMatrix}.
Predicate logic is leveraged for its interpretability and expressivity. 
Predicates are human-readable, supporting a user in understanding otherwise complex logic in machine learning and data analysis.
Predicates are also expressive -- a user can easily modify a predicate to express their domain knowledge.


More generally, a number of systems have used automated generation of interpretable rules to help users understand complex phenomena.
For example, by providing human-understandable summaries to better understand database provenance\cite{AlOmeir:2021:provenance}.
The Graphiti system follows a comparable approach, generating Boolean-logic rules to explain relationships between user-specified subsets from graphs by the user\cite{srinivasan2017graphiti}.
The DRIL system combines automatically and interactively generated rules to help analysts gain insights into the characteristics of interesting clusters of data points\cite{Cao2020DRIL}.

While these predicate or rule-based approaches have been applied to a number of common tasks in visual analytics, it has not yet been applied to understanding DR results.
DimBridge leverages these strengths of predicates to provide an interpretable \textit{bridge} between the projection and data-visualization spaces.

\section{Interpreting DR Results: Design Goal and Tasks}
Patterns identified in conventional DR visualizations are challenging to interpret because they lose the underlying semantics of the original dimensions, such as the context for why points are considered similar. 
Our work addresses these challenges by bridging the projection space and the underlying data space.
Specifically, \sys facilitates the interpretation of patterns found in DR results through interactive querying of the projection, enabling retrieval and visualization of relevant subspaces of the original dimensions.

We break down the process of identifying and interpreting patterns into three high-level tasks. These tasks are the foundational requirements used to develop the \sys system.

\subsubsection*{(T1) Identify and query visual patterns in a projection}
Dimensionality reduction (DR) transforms high-dimensional data into low-dimensional forms to facilitate visualization and analysis while preserving important structures and relationships.
A common approach is to reduce the data to two dimensions and visualize the resulting patterns using a scatterplot. 
\sys should support the user in identifying potential patterns of interest and facilitate direct engagement with the projection, enabling users to explore by selecting patterns that capture their interest. 
There are multiple patterns that a user can identify within a projection that have unique approaches to selection and interaction.
\sys must allow the user to identify and select each of the following patterns:

\smallskip\noindent \textbf{Clusters:} Identifying clusters is a key application of DR projections\cite{sacha:2016:dim_reduction_interaction_survey}, %
as it allows a way to classify similar and dissimilar points based on attributes. Although it can be intuitive to identify distinct groups of data points, it is not directly clear why they formed or how they differ by visualizing the projection alone. 

\smallskip\noindent \textbf{Outliers:} Identifying outliers is closely linked to recognizing clusters, yet determining if outliers in a 2D projection accurately reflect those in the high-dimensional data is challenging based solely on the projection. While outliers may be identified based on distance in the projection, it is not clear how this distance translates to the original dimensions.

\smallskip\noindent \textbf{Spatial Density:} Identifying clusters is not always straightforward, especially with varied point densities\cite{healey2011attention,szafir2016four}.
Unlike clear-cut clusters, density across a projection can change gradually, complicating the task of selecting a group of points for detailed analysis. For example, points just outside a selected dense area might still belong to that group, reflecting the challenge of making precise selections.



\smallskip\noindent \textbf{Shapes:}
DR projections are also used to uncover hidden low-dimensional structures in data\cite{sacha:2016:dim_reduction_interaction_survey}, such as identifying a low-dimensional manifold that explains the data distribution. These structures often appear as specific shapes within the projection, such as a curve slicing through a group of points.



\subsubsection*{(T2) Retrieve and visualize relevant subspaces of the original data}
Understanding visual patterns in the DR results requires explaining them in the context of the data's original dimensions.
Traditional visualization techniques, however, struggle with handling high-dimensional data effectively.
Given a visual pattern in the DR results queried by the users, \sys must be able to retrieve relevant subspaces of the original data.
For this subspace to be visualized effectively, it must have a relatively small number of dimensions.


 \subsubsection*{(T3) Evaluate the visual pattern in context}
While the goal is to explain a visual pattern in the original dimensions, distortions introduced by the DR algorithm can result in patterns in the DR results that do not correspond to patterns in the original dimensions.
Beyond simply retrieving relevant subspaces, \sys must help users evaluate a selected pattern given a retrieved subspace.
\section{Example Use Case and System Overview}
We provide a simple use case of \sys to illustrate how it addresses the design goals. 
Figure~\ref{fig:teaser} shows the example of a user with a set of animal images where each image is labeled with 14 numeric attributes (e.g., furry, whiskers, etc.).
After performing dimensionality reduction using UMAP and visualizing the results, the user identifies a cluster of data points that form an unusual curved shape in the \textit{Projection View}. 

The user investigates this pattern by selecting the points at the top of the cluster by drawing a bounding box, and then dragging the selected area to the bottom, following the cluster's curved shape (T1).
Given the selection, \sys identifies a set of (1) dimensions and (2) intervals for each dimension that most concisely explain the visual pattern (T2).
These dimensions are visualized using a scatterplot matrix in the \textit{SPLOM View}, with the corresponding intervals shown for each dimension in the \textit{Predicate View} (T3).
With these visualizations, the user can make the following observations:
\begin{itemize}[itemsep=2pt,parsep=3pt,leftmargin=*]
    \item The selected shape can be explained using just 4 of the original 14 dimensions.
    \item The shape most closely correlates with the ``furry'' and the ``whiskers'' dimensions.
    \item The top of the shape represents very furry animals with whiskers and big ears. Towards the bottom of the shape, the animals become larger, are less furry, and have small ears and no whiskers.
    \item There is a downward shift in the intervals for ``big ears'' where the shape curves to the right, corresponding to a tipping point where the animals' ears become increasingly smaller. 
    \item Confirming these observations, the user can see that data sampled along the shape represent animals from foxes to wolves, wolf-like dogs, large dogs, and finally smaller puppies.
\end{itemize}


To support this workflow, \sys models the user's interaction as a set of continuous brushes. In the example shown in Figure~\ref{fig:teaser}, the shape drawn by the user is discretized into 12 brushes, shown as 12 intervals for each dimension in the Predicate View.
To help the user make sense of this selection, \sys uses a novel predicate induction algorithm to identify a set of dimensions and intervals that best explain the brushes.
The algorithm enforces a \textbf{smoothness constraint}, such that the dimensions and the intervals in each of the predicates are consistent with the others.

In section \ref{sec:engine} we describe our predicate induction engine. Section \ref{sec:design} discusses the design of the visualization interface and interaction techniques. We demonstrate the utility of \sys with 3 examples and 2 use cases with collaborators in drug discovery and material science domains.

\section{Predicate Induction Engine}
\label{sec:engine}
The central motivation of this work is to support users' interpretion of patterns found in the DR results in the context of the original data dimensions.
The predicate induction engine facilitates this by generating predicates that best explain, in the original dimensions, a visual pattern identified by the user.




\subsection{Bridging DR and data patterns with predicates}
\sys uses first-order predicate logic as a ``bridge'' between patterns in the DR results and relevant subspaces of the original dimensions.
This allows users to capitalize on the strengths of both spaces: patterns can be identified visually in the DR results, and their semantics can be recovered in the original dimensions.

A first-order predicate, $\Phi$, is defined as a conjunction of one or more clauses, each consisting of a data dimension and a minimum and maximum value. The clause $\phi_{j}$ is defined for the $j$-th dimension, with minimum and maximum values denoted $\phi_{j}^{min}$ and $\phi_{j}^{max}$.
A predicate contains all data points with values that fall within the ranges defined by each clause.
Alternatively, a predicate can be thought of as a function that takes a data point, $x_{i}$, as input and outputs a binary label: 1 if the data point falls within the value ranges defined by each clause, and 0 otherwise:

\begin{align}
\Phi(x_{i}) = \mathds{1}[ \phi_{j}^{min} \leq x_{ij} \leq \phi_{j}^{max},\; \forall \phi_j \in \Phi]
\end{align}

\subsection{Generating predicates from user interactions}
The predicate induction engine enables \sys to leverage predicates as a bridge between patterns in the DR results and original dimensions by generating predicates that closely match the users selection.
First, we derive a set of background points, $B$, and pattern points, $P$, from a users interaction.
Second, we define the combined dataset $X = B \cup P$ and associated binary labels, $Y = \{\mathds{1}[x_{i} \in P] | x_{i} \in X\}$.
Finally, we use a predicate induction algorithm to identify predicates that contain a set of data points, represented by the binary labels $\Phi(X) = \{\Phi(x_{i}) | x_{i} \in X\}$, that closely match the data points selected by the user, represented by $Y$.

In our implementation of \sys, we consider two predicate induction algorithms.
The first is the recursive predicate induction (RPI) algorithm used by the PIXAL system\cite{montambault:2022:pixal}.
This algorithm constructs predicates from the ground up, starting with a set of single clause predicates that are iteratively refined. A predicate is scored at each step, and continuously refined until the score stops improving.
For \sys, we score predicates using the F1 score calculated between $\Phi(X)$ and $Y$, balancing false positives (points that are in the predicate but not the user's selection) and false negatives (points that are in the user's selection but not the predicate).
This results in multiple, overlapping predicates, each representing a candidate subspace.
While this may be suitable for offline applications, this is too slow for an interactive system.

The second is a novel algorithm, \alg that works by approximating high dimensional bounding cuboids via a differentiable proxy function.

\begin{figure}[!t]
    \centering
    {
    \includegraphics[width=\linewidth, trim={50px 200px 50px 250px},clip]{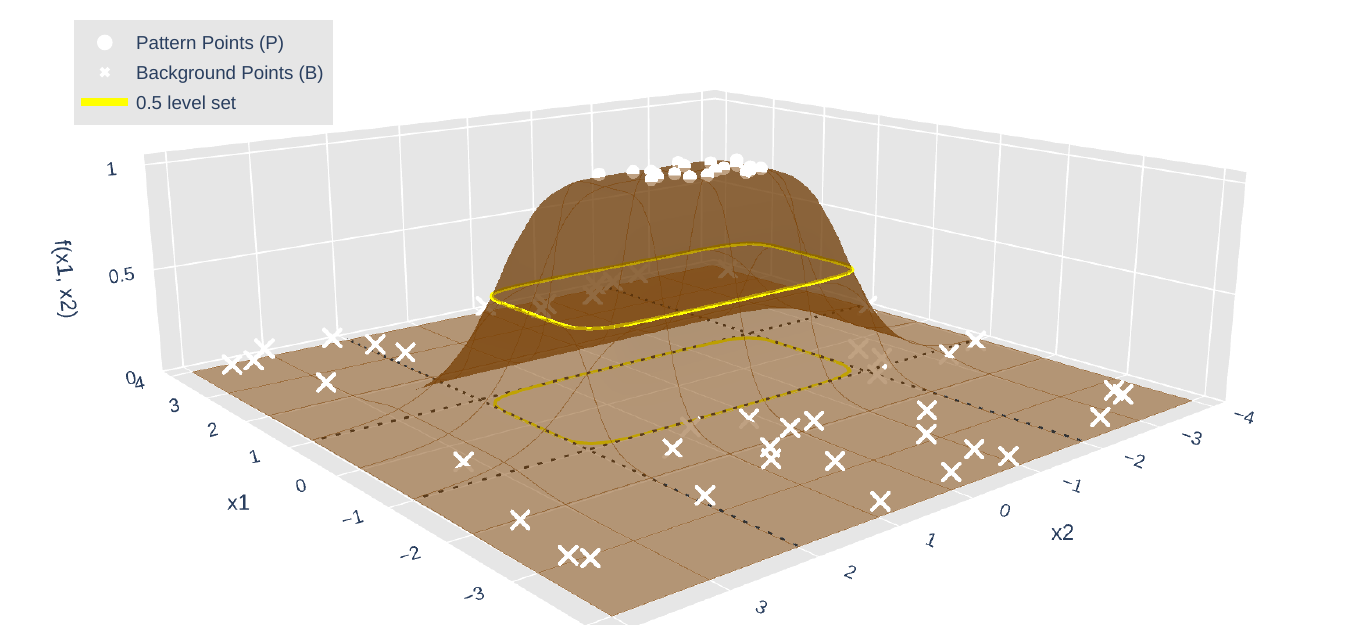}}%
    \caption{Illustration of differentiable proxy function (\autoref{eq:bump_function}) centered at $\boldsymbol{\mu} = (0,0)$ with $\mathbf{a}=(1, 1/2)$ and $b=7$.
    }%
    \label{fig:proxy-function}%
\end{figure}

\subsubsection{\alg}

Rather than generating and evaluating predicates iteratively, the \alg algorithm defines a differentiable loss function based on a reparameterization of first-order predicates, which it then minimizes to find a single ``best'' predicate given the $X$ and $Y$ defined by the user.




Note that when a predicate clause fully cover the data extent along certain dimension, the clause always return true. 
Therefore, even though a predicate defines a subspace that only includes dimensions in the clause, we can expand it to all dimensions, with the range being strictly greater than the data extent:

\begin{align}
\Phi(x_{i}) = \mathds{1}[ \phi_{j}^{min} \leq x_{ij} \leq \phi_{j}^{max},\; \forall j=1 \ldots M]
\end{align}
with $\phi_j^{max} \geq \max\limits_{i}\{x_{ij}\}$
and  $\phi_j^{min} \leq \min\limits_{i}\{x_{ij}\}$
when  $\phi_j \notin \Phi$.

This expansion gives us opportunity to optimize all predicate clauses simultaneously.
Moreover, we can define a predicate with the parameters $\boldsymbol{\mu}$ and $\mathbf{r}$, denoting a midpoint and range for each dimension ($j$) respectively:


\begin{align}
\mu_j = \frac{\phi_{j}^{max} + \phi_{j}^{min}}{2} 
\text{, }\;
r_j = \frac{\phi_{j}^{max} - \phi_{j}^{min}}{2} 
\end{align}



This reparameterization allows us to consider whether a data point is contained by a predicate not only as a binary label, but a continuous probability:

\begin{align} 
Pr(\Phi(x_{i})=1 | \mathbf{r}, \boldsymbol{\mu}, b) := \frac{1}{1 + \sum_{j=1}^{M} | \frac{1}{r_{j}} \cdot (x_{ij} - \mu_{j})|^{b}} 
\end{align}

\paragraph{Generating a Predicate for a Single Brush:} Geometrically, the probability gives a rounded bump function (see \autoref{fig:proxy-function}) of $x_i$ centering at $\boldsymbol{\mu}$, where
$b$ is a fixed parameter controlling the steepness of the bump. 
Moreover, the $0.5$ level set is enclosed by the predicate bounding box, $\prod_{j=1}^{M}[\mu_j - r_j, \mu_j + r_j]$, along each data feature dimension $j$. 
To find the optimal predicate via optimizing the parameters, we rewrite $1/r_j$ as $a_j$ and view the probability as a differentiable function of $a_j$ and $\mu_j$:
\begin{align}
f(x_i | \mathbf{a}, \boldsymbol{\mu}, b)
= \frac{1}{1+ \sum_{j=1}^{M} |a_j \cdot (x_{ij} - \mu_{j})|^{b}} 
\label{eq:bump_function}
\end{align}


Given an $X$ and $Y$ defined by a user's selection, the loss function, binary cross entropy (BCE), is defined as:


\begin{align*}
\mathcal{L}_{bce}(\mathbf{a}, \boldsymbol{\mu} | X, Y) 
= \frac{1}{N} \sum_{i=1}^{N} y_{i}\log(f(x_{i})) + (1 - y_{i})\log(1 - f(x_{i}))
\end{align*}

Recall that $a_j$ is the inverse of the range $r_j$, and enlarging the range beyond data extent (i.e. forcing $a_j \to 0$) effectively eliminates the corresponding clause in the predicate conjunction.
Therefore, selection of features in the predicate can be achieved through an $L_1$ regularization, $||\mathbf{a}||_1$.
Eventually, minimizing loss functions over $\mathbf{a}$ and $\boldsymbol{\mu}$ gives a predicate in the form $\prod_{j=1}^{M}[\mu^*_j - 1/a^*_j, \mu^*_j + 1/a^*_j]$, where

\begin{align}
\mathbf{a}^*, \boldsymbol{\mu}^* =  
\argmin_{\mathbf{a}, \boldsymbol{\mu}} 
\mathcal{L}_{bce}(\mathbf{a}, \boldsymbol{\mu}) + \gamma_1 \cdot ||\mathbf{a}||_1
\end{align}
and $\gamma_1$ controls the strength of the predicate sparsity.

\paragraph{Generating Predicates for Multiple Continuous Brushes:}
One requirement of a predicate induction algorithm, when it comes to the interactive nature of \sys, is the consistency and smoothness of results it gives from consecutive interactions.
For example, one may expect smoothly changing clauses when fine-tuning queries, or brushing curves over a region.
For these cases, a constraint can be added to the objective function that encourages smoothness in the resulting predicates.

The draw interaction results in a discretized sequence of selections derived from the user's gesture.
Given a sequence $X_t, Y_t$ for $t = 1...T$, optimizing
\begin{align}
\sum_{t=1}^{T} \mathcal{L}_{bce}(\mathbf{a_t}, \boldsymbol{\mu_t} | X_t, Y_t)
\end{align}
gives a sequence of predicates independent to one another.
To encourage consistency and continuity between consecutive predicates, a smoothness loss function is introduced:
\begin{align}
\mathcal{L}_{smooth} = \sum_{t=2}^{T} 
\gamma_{a} \cdot ||\mathbf{a_t} - \mathbf{a_{t-1}}||^2
+ \gamma_{\mu} \cdot ||\boldsymbol{\mu_{t}} - \boldsymbol{\mu_{t-1}}||^2
\end{align}
where $\gamma_{a}$ and $\gamma_{\mu}$ control the strength of consistency and continuity between consecutive predicates.

In entirety, the loss function becomes
\begin{align}
\mathcal{L} =& \sum_{t=1}^{T} \mathcal{L}_{bce}(\mathbf{a_t}, \boldsymbol{\mu_t} | X_t, Y_t) 
+\sum_{t=1}^{T} \gamma_{1} \cdot ||\mathbf{a_t} ||_1 \\
&+ \sum_{t=2}^{T} \gamma_{a} \cdot ||\mathbf{a_t} - \mathbf{a_{t-1}}||^2 
+ \sum_{t=2}^{T} \gamma_{\mu} \cdot ||\boldsymbol{\mu_t} - \boldsymbol{\mu_{t-1}}||^2\ 
\end{align}

\section{Visual Interface}
\label{sec:design}

The \sys interface consists of three coordinated views: The \textit{Projection View}, the \textit{Predicate View}, and the \textit{SPLOM View}.
By interacting with these views, users can query patterns in the DR results, view and modify predicates, visualize relevant subspaces of the original dimensions, and evaluate the queried pattern in context.

Users can employ the Projection View to visualize and interact with patterns through a scatterplot displaying the DR results.
The Predicate View allows users to view and adjust predicates generated by the Predicate Induction Engine in response to these interactions.
Finally, the SPLOM View allows users to visualize the data in a subspace of the original dimensions defined by these predicates.

Predicates are generated with the goal of matching a user's selection as closely as possible. However, distortions introduced by the DR algorithm will results in some data points that are in the user's selection not being included in the predicate, and some data points in the predicate not being in the user's selection.
These mismatches are illustrated using color in the Projection and SPLOM Views.
Similar to prior work\cite{jeon2022distortion}, this provides \sys users with a transparent indication of whether a perceived cluster can be adequately explained by a first-order predicate in the original dimensions.






\subsection{Visualizing Data}
In the Projection View, the DR results are visualized with a scatterplot.
A user can interact directly with the scatterplot to identify interesting patterns by brushing the related data points.
Brushing in the Projection View is supported by two basic interactions. Users can \textit{select} a region of the projection by drawing a bounding box or lasso around that region, or \textit{draw} a selected region between one location in the scatterplot and another.
These two brushing interactions allow \sys users to select a wide variety of interesting visual patterns that might appear in the DR results:

\smallskip\noindent \textbf{Select}:
A user can select a group of data points ($P$) that are distinct from the rest of the dataset ($B$) using the bounding box or lasso.
In response, the Predicate Induction Engine will attempt to generate predicates that distinguish $P$ from $B$ in the original dimensions.
This can be used to identify \textbf{clusters} that are unique relative to the rest of the data due to their shape or positioning.
Similarly, this can be used to identify groups of global \textbf{outliers} that are spatially distant from the rest of the data, as well as regions with lower or higher \textbf{spatial density}.

\smallskip\noindent \textbf{Select and Contrast}:
Rather than comparing to the rest of the data, a user can use the bounding box or lasso to select another group of data points ($B$) to compare to.
This interaction is used to explain the difference between two \textbf{clusters}, local \textbf{outliers} and their neighbors, or to explain variation in \textbf{spatial density} relative to a local region.

\smallskip\noindent \textbf{Select and Draw}:
If a group of data points forms a distinct, continuous \textbf{shape}, a user can use the bounding box or lasso to select a group of data points to define the shape's starting point. Then, they can drag the selected region to the shape's endpoint, drawing through the rest of the data points along the way.
In response, the Predicate Induction Engine will return a series of predicates that distinguish the selected points ($P$) at multiple intervals along the shape.

In the SPLOM View, a subspace of the original dimensions is visualized with a scatterplot matrix.
Rather than visualizing every dimension, the SPLOM View includes only those dimensions that are included in a predicate.
This reduces clutter in the visualization and helps a user focus only on the dimensions of relevance.
\subsection{Visualizing Predicates}


In the Predicate View, a user can visualize and interact with predicates generated by the induction engine.
This view lists each dimension included in a predicate and displays its value ranges.
Value ranges have multiple possible representations, depending on the user's interaction:

\smallskip\noindent \textbf{Select:} Each dimension is displayed next to a horizontal bar whose total length represents its full range of values.
A segment of each bar is highlighted, representing the range of values defined by the predicate (Fig. \ref{fig:cheetah}).
Users can directly modify the predicate by clicking and dragging either endpoint to adjust the range of the predicate (Fig.~\ref{fig:showcase:diabetes}, \ref{fig:case_study_uninteresting}).

\smallskip\noindent \textbf{Select and Contrast:}
Instead of one value range, two ranges (one for each selection) are displayed for each dimension (Fig.~\ref{fig:cat-dog-contrast}).

\smallskip\noindent \textbf{Select and Draw:}
A user's draw gesture is discretized into a set of discrete selections. Multiple value ranges (one for each selection) are displayed vertically to preserve space (Fig.~\ref{fig:teaser}).


\smallskip
The set of data points that fall within the predicate are highlighted in both the Projection View and the SPLOM View.
Situated between the Projection View and the SPLOM View, the Predicate View acts as a conceptual bridge between a pattern in the DR results and patterns in the original dimensions.
Users can modify a predicate by adjusting the value ranges or adding or removing a dimension to observe how these changes impact the Projection and SPLOM Views.


Points in the Projection and SPLOM Views are categorized and assigned a color based on this predicate.
Points within the user's selection and the predicate (true positives) are shown in purple, those in the predicate but not the selection (false positives) are shown in red, and those in the selection but not the predicate (false negatives) are shown in blue. All other points (true negatives) are shown in grey.
Categories update dynamically in response to interactions in the Projection (updating the selection) and Predicate (updating the predicate) Views.

\section{Showcases}
\label{sec:showcases}


In this section, we provide three showcases using \sys to analyze image data, motion-capture data, and scientific data.
Each of these showcases highlights a user's task in exploring and making sense of high-dimensional data.

\subsection{Understanding the Space of a Generative Model}
As generative AI becomes more commonplace in today's society, understanding a generative model's affordances is vital for instilling model trust\cite{vasconcelos2023generation,friedrich2023fair}.
We employ \sys for analyzing the output space of a generative vision model to demonstrate the range of patterns one encounters in DR projections and to illustrate how \sys provides the necessary context for making sense of these patterns.
More specifically, we consider the StyleGAN3\cite{karras2021alias} image generation model, which aims to generate an image collection depicting animals.
We define a collection of textual phrases as attributes -- 14 in total -- that describe visual appearances one would expect of animals, e.g. ``furry'', ``size'', and descriptions of behavior, e.g. ``excited'', or ``suspicious looking''.
For each phrase, we use CLIP\cite{radford2021learning} to score the attribute against a given generated image. We take this set of continuously-valued attributes as our high-dimensional space alongside the generated images to verify findings.
\begin{figure}[!t]
    \centering
    {\includegraphics[width=\linewidth]{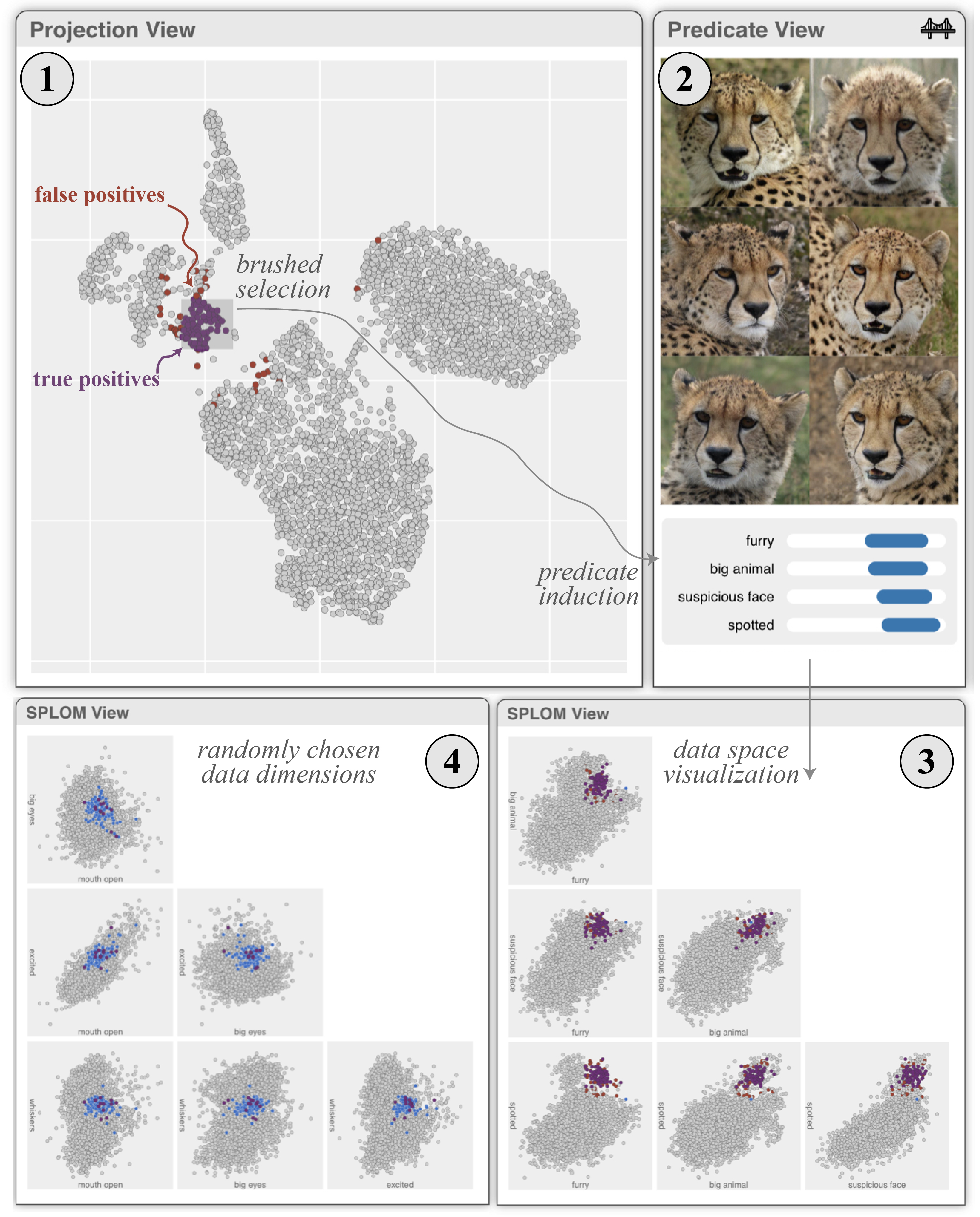}}
    \caption{
        We show how \sys allows one to better understand a potential cluster within the output space of a generative vision model.
        Upon performing a brush in the scatterplot (1), \sys finds a predicate comprised of 4 attributes (2) that, combined, help distinguish cheetahs from other animals (3), e.g. a big animal with spotted features.
        In comparison, highlighting brushed data points in randomly chosen four features (4) does not help in distinguishing key features of cheetahs from other animals.
        }%
    \label{fig:cheetah}%
\end{figure}
\begin{figure}[!t]
    \centering
    {\includegraphics[width=\linewidth]{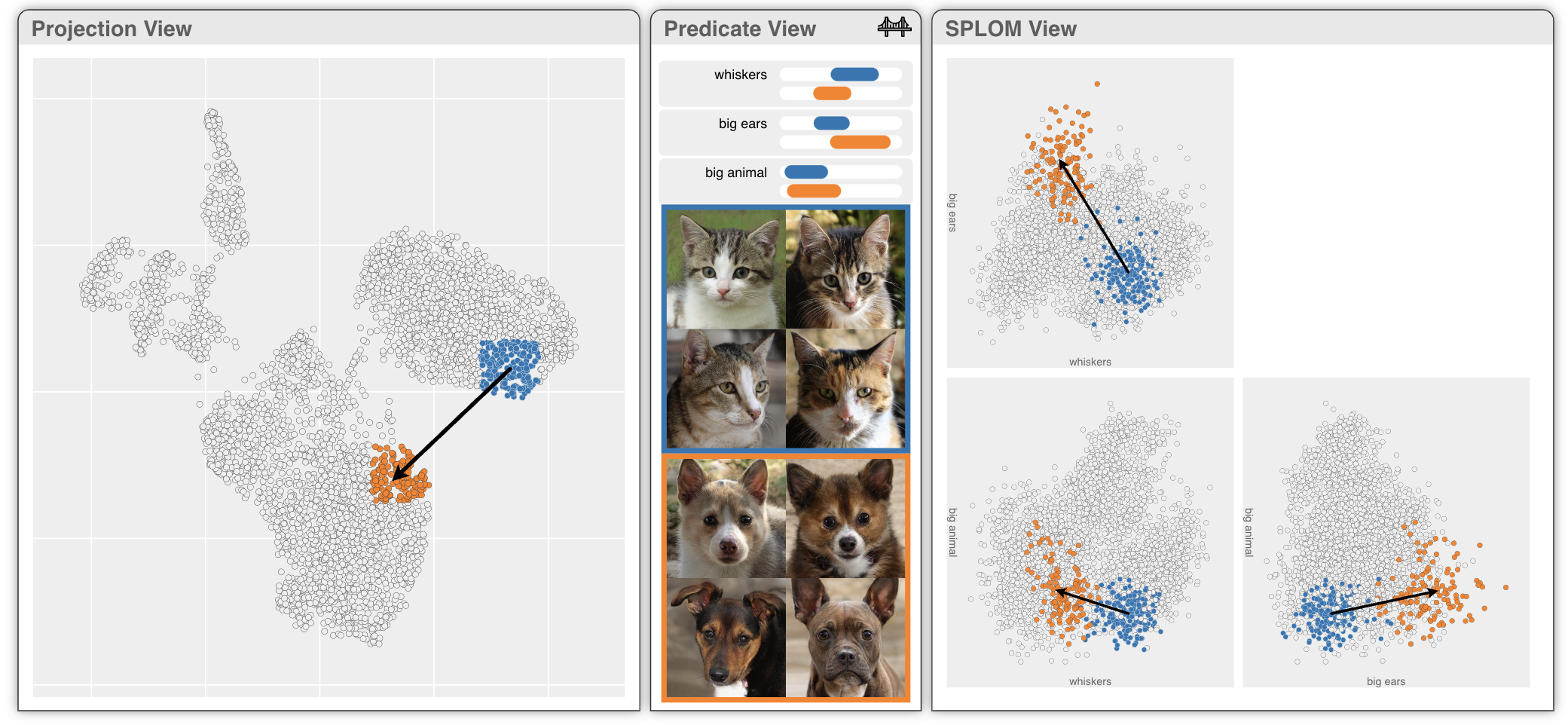}}%
    \caption{
        We show how \sys allows one to better contrast one region of the dimensionality reduction plot from another. Upon brushing two regions, \sys finds a predicate that explains the two regions from the rest of the data points.
        \sys finds that while both kittens (blue) and puppies (orange) are not big animals, kittens have whiskers, and puppies have bigger ears.
        }%
    \label{fig:cat-dog-contrast}%
\end{figure}
%

\smallskip
\noindent \textbf{Explaining a Cluster:} 
\sys demonstrates cluster analysis in a UMAP projection, highlighting discernible groups within the data, as seen in the scatterplot (Fig.~\ref{fig:cheetah}).
Upon brushing one of the clusters, \sys derives the predicate that best distinguishes those data items from the rest of the dataset.
The predicate reveals the data items represent large, spotted animals with distinct facial expressions, identified as cheetahs upon reviewing selected images. A scatterplot matrix (Fig.~\ref{fig:cheetah}.3-4) contextualizes this, showing cheetahs as uniquely furry and spotted within the dataset.
We also find that there are a small number of data items that are false positives (colored in red), e.g. animals that fall within the predicate.
Some of these data items are close to the brushed region and we observe these are mostly cheetahs, indicating that the perceived cluster that was brushed was slightly imprecise.

As a baseline for comparison, we chose four dimensions and plotted the user's selection in the SPLOM (Fig.~\ref{fig:cheetah}-4).
Plotting data on these dimensions hardly contextualizes the user's selection against the dataset, whereas the algorithmically defined
predicates allow one to reason about \emph{why} this cluster is discriminative, highlighted in the scatterplot views (Fig.~\ref{fig:cheetah}-3).

\smallskip
\noindent \textbf{Contrasting Two Regions in Context:}
Beyond comparing a cluster with the overall dataset, users can also compare one cluster to another. 
In Fig.~\ref{fig:cat-dog-contrast}, \sys highlights the differences between kitten and puppy clusters in the DR projection, distinguishing them by features such as "whiskers" and "big ears" and noting they are both categorized as smaller than other animals in the dataset.

\subsection{Understanding Progression in Motion Captures}

\begin{figure}[!t]
    \centering
    {\includegraphics[width=0.33\linewidth]{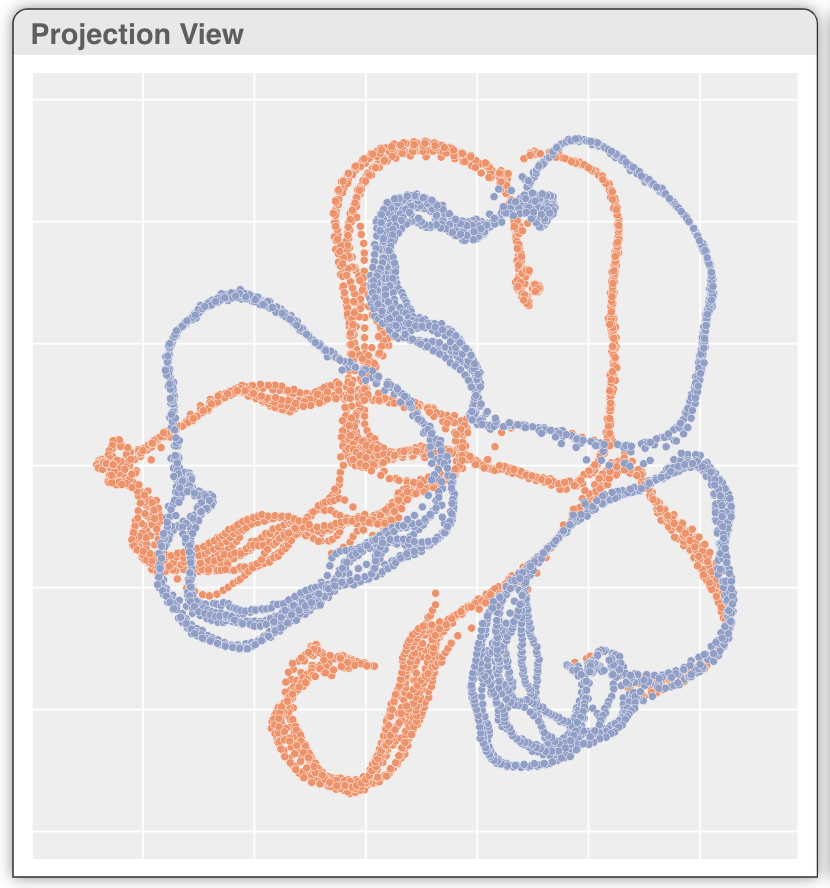}}%
    {\includegraphics[width=0.33\linewidth]{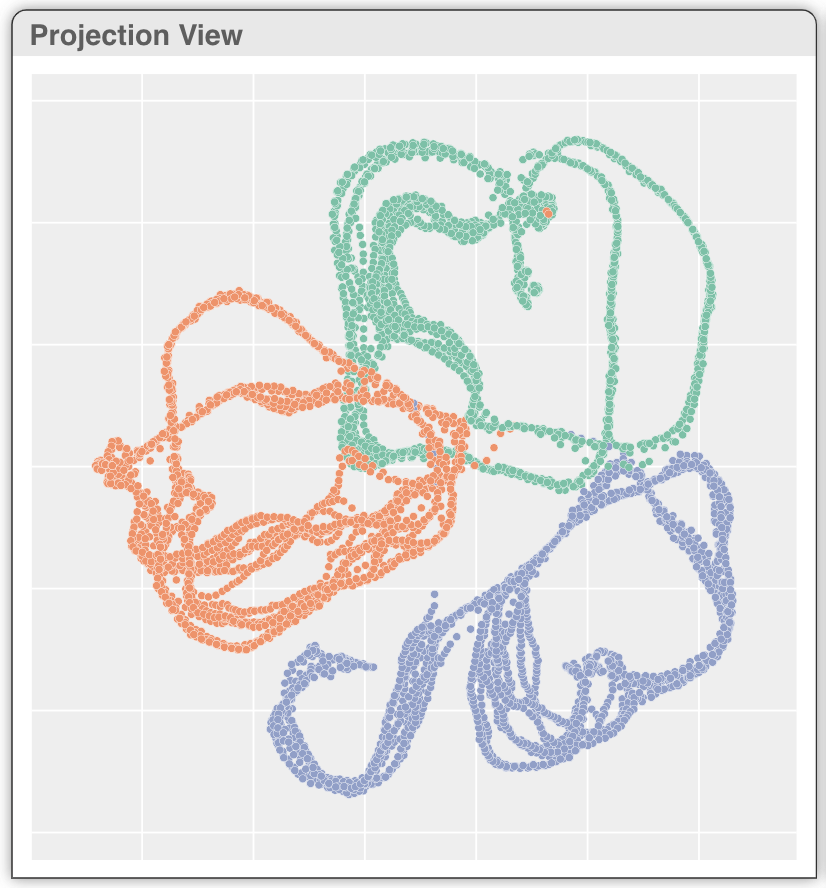}}%
    {\includegraphics[width=0.33\linewidth]{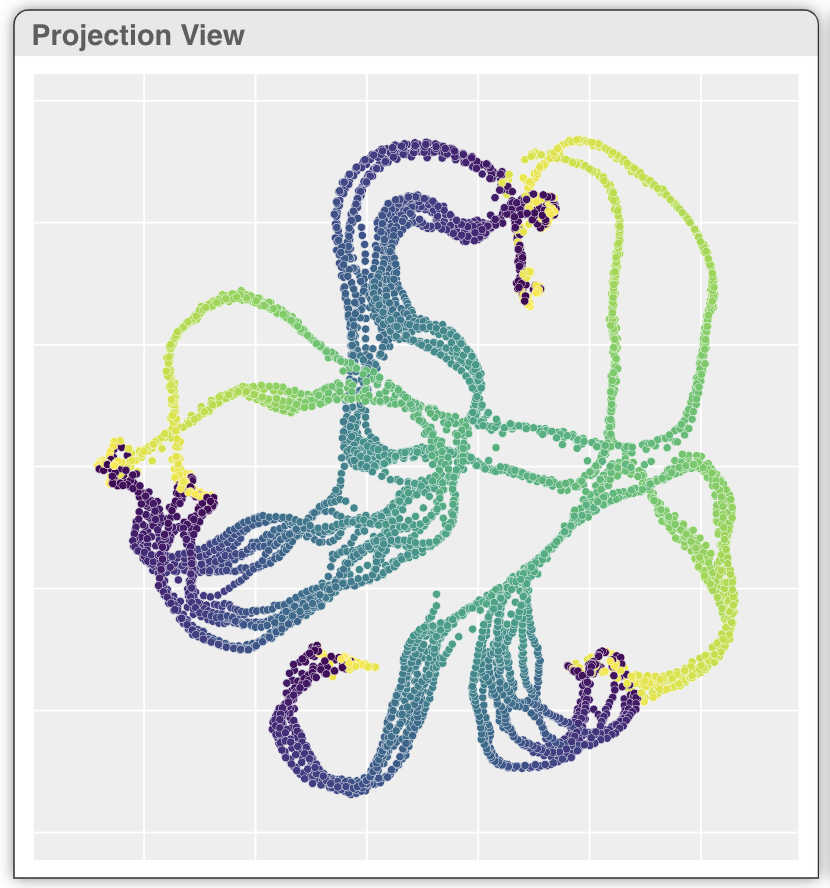}}%
    \caption{DR plot of the Motion Capture dataset. \textbf{Left:} color by subject. \textbf{Middle:} color by bracing conditions. \textbf{Right:} color by time.}%
    \label{fig:mocap-ground-truth}%
\end{figure}

Biomechanical data, such as cyclic motion recordings, are crucial in orthopedic, rehabilitation, and sports research\cite{helwig2016smoothing}. 
The Multivariate Gait Dataset\cite{misc_multivariate_gait_data_760, shorter2008new, helwig2011methods, helwig2016smoothing} captures the motion of walking humans in terms of 6 joints angles (\{left, right\} $\times$ \{ankle, knee, hip\}) for 101 timestamps and 10 repetitions, among 10 subjects and under 3 bracing conditions (\{unbraced, knee brace, and ankle brace\}).
Here we selected 2 subjects from the 10 in the original dataset for illustration.
The dimensionality reduction plot is generated via UMAP using only the 6 angular features, excluding the explicit encoding of timestamps, repetitions, subject IDs, and bracing conditions.
In the DR projection scatterplot (Fig. \ref{fig:mocap-ground-truth}), we identified three groups of loops corresponding to the three bracing conditions, each containing two overlapping repetition bundles.
The difference between the two bundles comes from the subject, as shown in Fig.~\ref{fig:mocap-ground-truth}.
Coloring the DR plot by attributes helps understand data one attribute at a time, but simultaneously analyzing multiple continuous attributes can be challenging. In a motion capture dataset, displaying all 6 joint angles in a SPLOM creates a cognitive load due to the $\binom{6}{2}$ possible subplots. However, \sys generates predicates that highlight a few key views, greatly easing the visual workload in exploratory data analysis.

\begin{figure}[!t]
    \centering
    {\includegraphics[width=\linewidth]{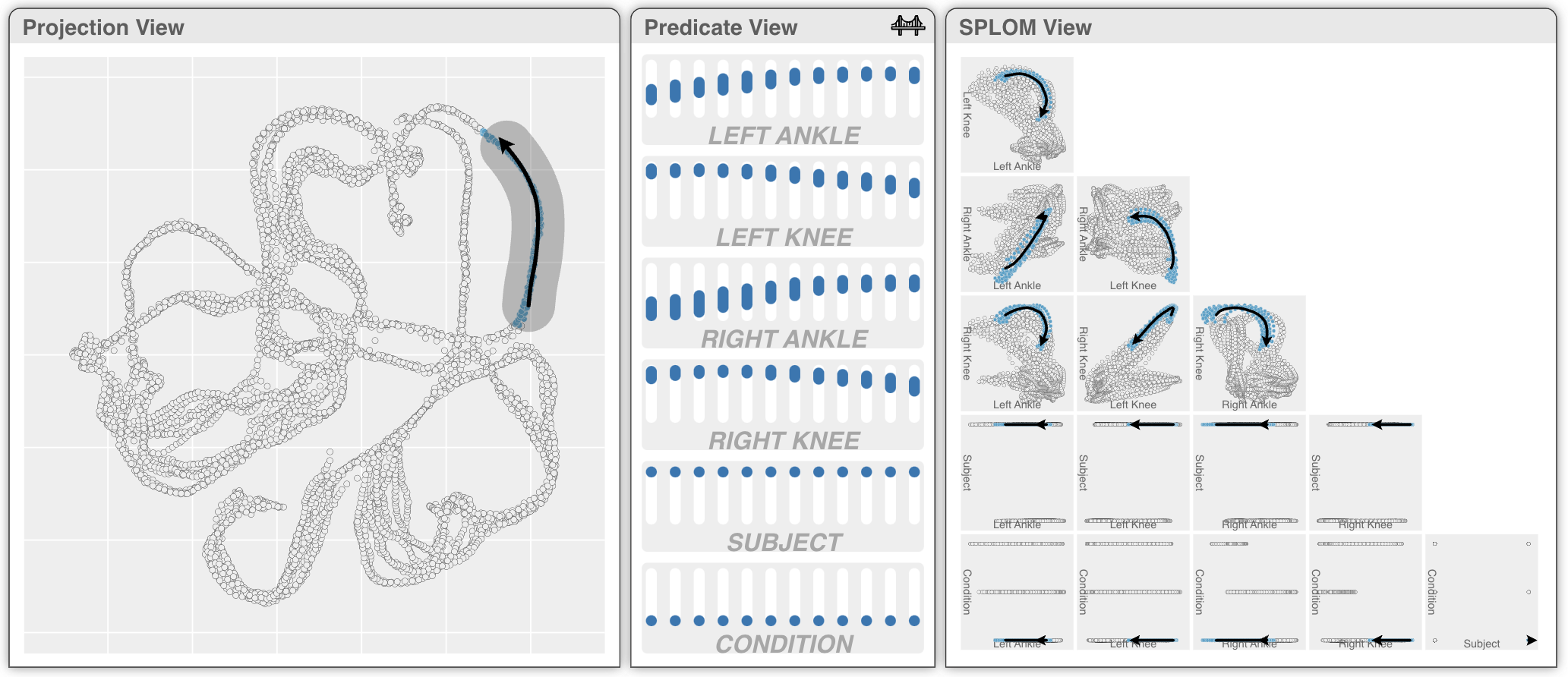}}%
    \caption{\sys shows that the curve following the flow of time in the figure captures only one subject and condition, and the segment represents a period with increased angles on left and right ankles and slightly decreased angles on left and right knees.}%
    \label{fig:mocap-curve}%
\end{figure}

\smallskip
\noindent \textbf{Understanding Progression Over a Curve:}
Brushing over a segment of one group in the direction of time flow, the predicate induction algorithm recognizes that the segment belongs to a single subject under a particular bracing condition (Fig.~\ref{fig:mocap-curve}).
It further summarizes the progression, in the direction pointed by arrows on the plot, as an increase in left and right ankle angles with a slight decrease in angles on left and right knees.
The SPLOM View confirms that this summary indeed distinguishes the brushed segment from the rest of data items.

%
%
\subsection{Examining Populations within a Diabetes Study}
\label{sec:showcases:diabetes}
\begin{figure}[!t]
    \centering
    \includegraphics[width=\linewidth]{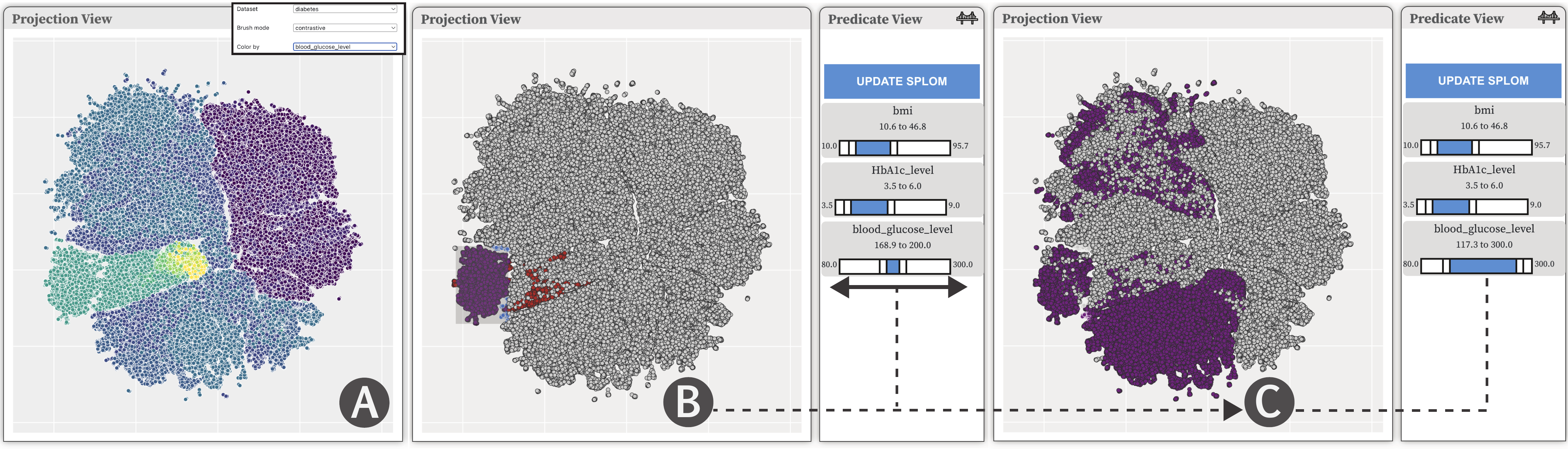}
    \caption{
A system screenshot showing the projection and predicate components: (A) Users start by coloring the projection according to a feature, here Blood Glucose level. (B) A domain practitioner explores a data subset, creating a predicate based on their selection. (C) They then adjust the Blood Glucose range to a meaningful one, observing changes in the point distribution.}
    \label{fig:showcase:diabetes}
\end{figure}
Healthcare remains one of the fastest-growing industries in the U.S.\cite{BLS:2023:HealthLaborStats}.
The industry is rapidly modernizing, and producing data at an ever-increasing rate\cite{Ristevski:2018:BigDataInHealthcare}.
There is a need for new techniques that can help make sense of this influx of data.

In this use case, we examine a dataset focused on diabetes.
We examine nine features, as well as a quantitative measure of how far the disease has progressed following a year.
The nine features are age, body mass index, blood pressure, total serum cholesterol, low-density lipoproteins,  high-density lipoproteins, total cholesterol,  serum triglycerides level, and blood sugar level.

\smallskip
\noindent \textbf{Specifying a Known Population}: We demonstrate how a practitioner with domain knowledge can use \sys to manually adjust ranges within returned predicates to find specific populations within the projection.
The user begins by adding relevant features to those used in the projection and
SPLOM in the data-space visualization. 
In this case, the user is interested in BMI, hbA1c level, blood glucose level, and age. 

As seen in Fig.~\ref{fig:showcase:diabetes}-A, they can color the projection by values for blood glucose values, finding a subset of the data of interest in the lower left of the projection. They can then generate initial predicates from the brush selection they have made to the projection (Fig.~\ref{fig:showcase:diabetes}-B).

\smallskip
\noindent \textbf{Interactive Predicate Refinement:}
The user can then adjust the blood glucose level ranges in the predicate view to focus on particular states of the condition (Fig.\ref{fig:showcase:diabetes}-B). By manipulating these ranges, the system dynamically updates the projection view, highlighting the data points that fall within the newly specified thresholds. This process is visualized in Fig.\ref{fig:showcase:diabetes}-C, where the adjusted range causes a redistribution of the highlighted points, providing immediate visual feedback on the change.

The ability to update these ranges is not only important for identifying distinct diabetic populations but also for exploring the complex interplay between glucose levels and other biomarkers. Upon updating the ranges to the desired distribution, the SPLOM can also be updated by clicking the "Update SPLOM" button, further allowing users to examine the pairwise relationships between the blood glucose levels and other variables such as BMI, HbA1c, and age. This refined analysis can surface more subtle correlations or patterns, which might go unnoticed with a more static approach.


\section{Case Study: Investigating Properties of the Mn1-xGexTe Alloy}
\label{sec:casestudies}
The case study presents an application of \sys to materials science, focusing on research to better understand the local structure of non-periodic materials such as alloys, that are difficult to model with traditional methods. Stoichiometric analysis in materials science often relies on determining the proportions of elements and their interactions. For the case study in question, our collaborators used sophisticated sampling techniques recommended by Novick et al. \cite{novick2023simulating}, along with density functional theory (DFT) — a quantum mechanical method that calculates the properties of materials based on electron behavior — to explore the structure of the Mn1-xGexTe alloy as it changes with the chemistry. 
This process created around 500 theoretical model representations, to explore the material's structure with different combinations of Manganese (Mn) and Germanium (Ge). These models are grouped into five types based on their Mn to Ge ratios:
\ce{Mn_{0.125}Ge_{0.875}Te}, \ce{Mn_{0.2}Ge_{0.8}Te},
\ce{Mn_{0.25}Ge_{0.75}Te},
\ce{Mn_{0.3}Ge_{0.7}Te},
\ce{Mn_{0.375}Ge_{0.625}Te}.

\subsection{Theoretical Models and Empirical Validation: Current Analysis Methods}
Characterizing theoretical models is crucial for understanding how atoms bond and arrange themselves and the local structure, which directly relates to the properties of the material. 
For non-periodic materials like alloys, characterization involves theoretical and empirical methods to simulate atom distributions. 
Standard analysis involves comparing these detailed computational predictions with actual experimental data.

For the data used in this case study, the real-world empirical data used to confirm the theoretical models were obtained from the Spallation Neutron Source at Oak Ridge National Lab
Researchers traditionally visualize and analyze data through computational simulations, graphical representations, and empirical data comparisons. In this area of research, Pair Distribution Functions (PDF) plots\cite{young2011applications} are used to understand and compare the local atomic structure of different alloy compositions to one another and the experimental data
Signals identified in the data are referred to as ``peaks'', which we will refer to in the subsequent sections. Generating lower-dimensional representations of the data is straightforward, but connecting these to the underlying science driving the clustering is very difficult with our collaborators’ current workflow.

\subsection{Analyzing the Mn1-xGexTe Alloy with Dimbridge}
Our collaborators were interested in using \sys as an exploratory analysis tool to understand the similar characteristics shared by the computationally generated theoretical models for the Mn1-xGexTe alloy that
align with the experimental data from Oak Ridge National Lab. Specifically, they asked the following questions: \textbf{(1)} How does changing the proportion of manganese (Mn) affect the way the alloy's structure shifts from a slanted, diamond-like shape to a straight-edged, cube-like shape? \textbf{(2)} Are the shapes formed by Mn atoms similar across the alloy space? For example, if we look at shapes with 6 out of 12 manganese atoms in their outer layer, do they look the same from one Mn-Ge mixture to another?


\subsection{Walkthrough of Analysis with Dimbridge}
During our collaborator's use of \sys, it became apparent that the system is not only a tool for bridging dimensional spaces but a catalyst for testing hypotheses and identifying areas for rich exploration.

Our collaborator for this case study, a graduate student at an affiliate university, studies how varying the ratio of Mn and Ge in alloy samples induces a transition between crystal structures, to better understand the relationship between local atomic coordination, global crystal structure, and material properties. Having a deep familiarity with the data, she familiarized herself with \sys through a process of generating hypotheses about clusters, finding an area of interest, and refining both her mental model of the data and investigating strategies for subsequent data runs.

Her exploration and analysis with \sys progressed in three distinct iterations, defined by a revised dataset tailored to a new investigative strategy.
All three of these iterations used datasets generated from the five subsets but differed in the properties for these subsets.
As mentioned previously, subsets varied in Mn to Ge combinations: 
\ce{Mn_{0.125}Ge_{0.875}Te}, \ce{Mn_{0.2}Ge_{0.8}Te},
\ce{Mn_{0.25}Ge_{0.75}Te},
\ce{Mn_{0.3}Ge_{0.7}Te},
\ce{Mn_{0.375}Ge_{0.625}Te}.
 
\begin{figure}[!h]
    \centering
    {\includegraphics[width=\linewidth]{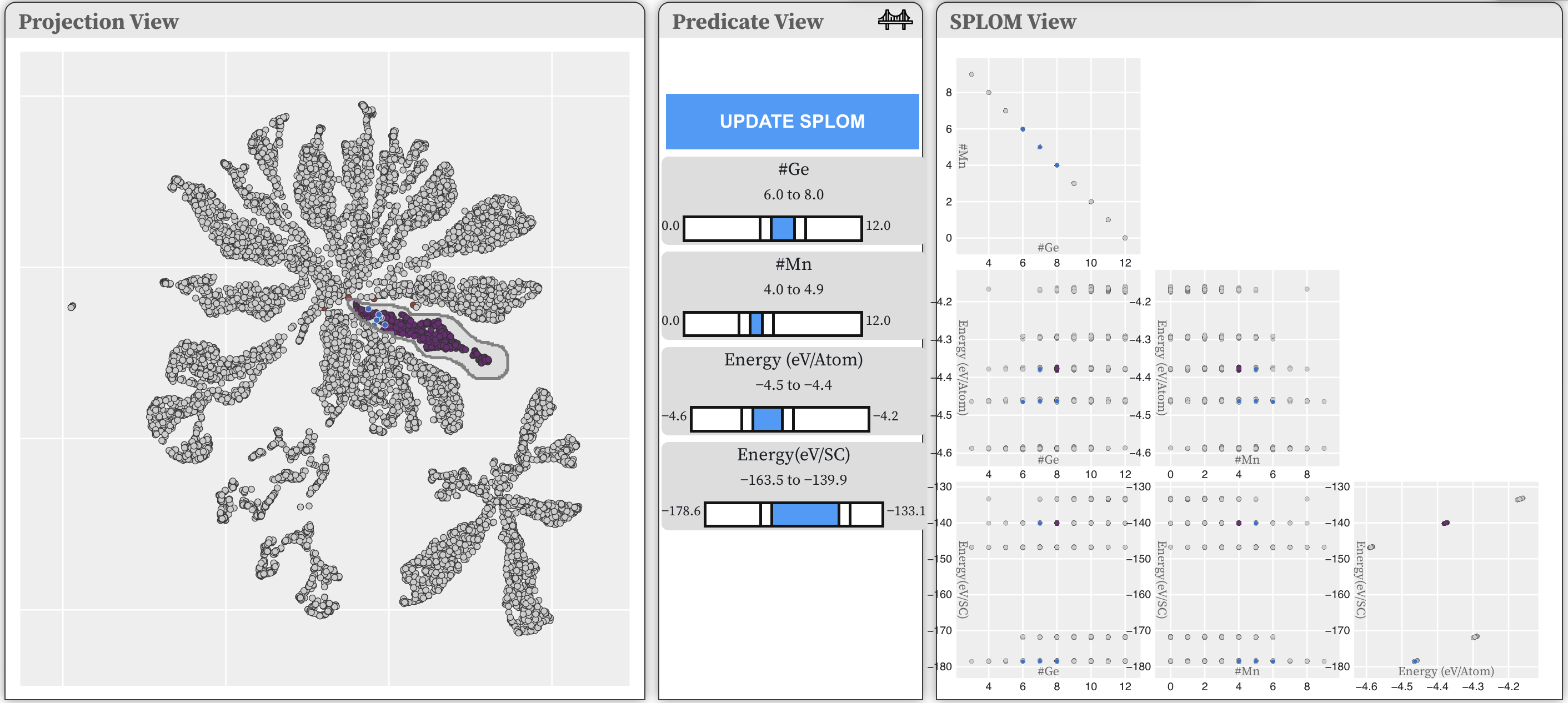}}
    \caption{UI of \sys with cluster selected, showing the division of points into clean groups in the SPLOM. This indicates to our collaborator the strong influence Mn has on the clustering of data.}%
    \label{fig:case_study_uninteresting}%
\end{figure}
\subsubsection{Phase one: Investigating Mn's Influence on Separation}

The first phase of exploration involved a broad exploration of all five subsets in which she experimented with various hyperparameters to investigate their impact on data separation, including adjusting and observing the effects of removing attributes with predictable influences on the dataset's division into clusters, specifically investigate whether the Mn coordination shell was important.

Between the removal of attributes, the projection would be color-coded again to confirm separations she had predicated would be present, such as the division of each subset into largely disparate clusters.
She would then brush clusters of interest to inspect any correlations between attributes of interest or confirm that the clusters she was inspecting resulted from something uninteresting, such as the SPLOM in Figure \ref{fig:case_study_uninteresting}, which were so separated by the proportion of Mn and Ge, it generated results uninformative to our collaborator.

\begin{figure}[!t]
    \centering
    \includegraphics[width=\linewidth]{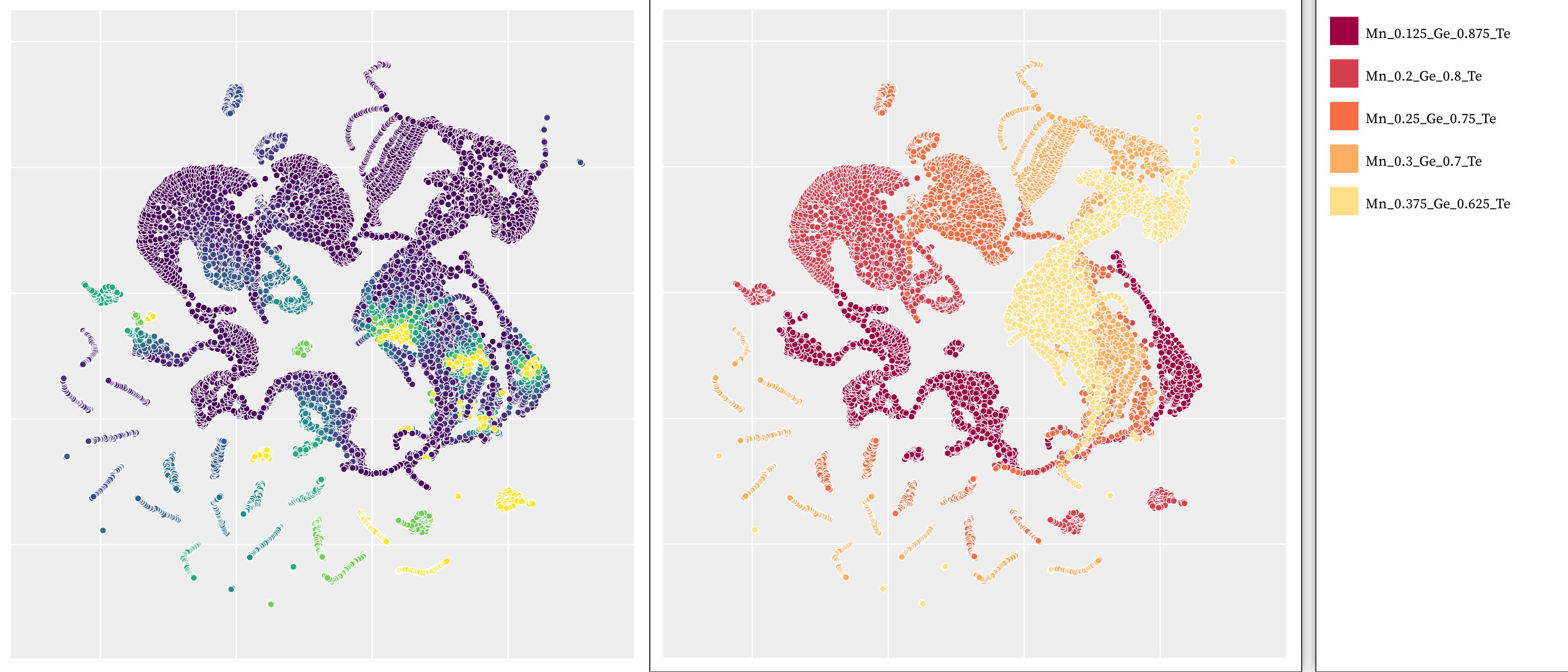}
    \caption{Projection view showing temperature dependencies color-coded by attributes. The left is color-coded by temperature and the right is color-coded by the data subset indicating the varying combinations of Mn and Ge.
    }
    \label{fig:case:large_projection_temp_color}
\end{figure}

\subsubsection{Phase two: Investigating Temperature Dependencies}
\begin{figure}[h!]
    \centering
    {\includegraphics[width=\linewidth]{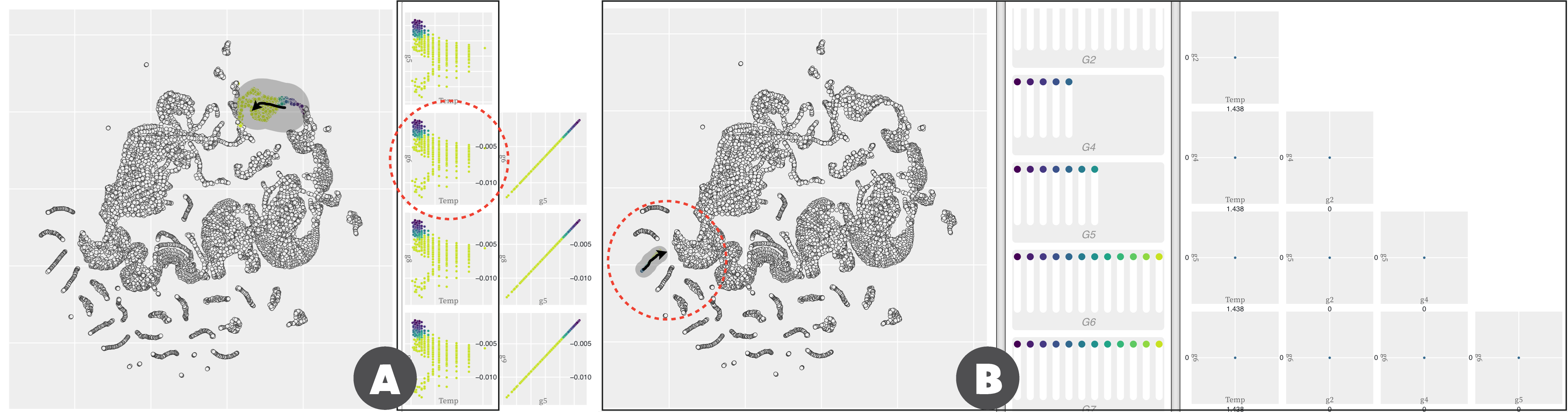}}
    \caption{(A) \sys explains a trail within the alloy temperature prediction dataset. The SPLOM in the data view shows curves merging as the temperature increases. (B) View of the data subset of the drawn shape selection, indicating that the cluster selection contains only one temperature value.}%
    \label{fig:case_study:merging_curve}%
\end{figure}

Her findings indicated that the number of Mn atoms significantly influenced data separation in lower-dimensional spaces, prompting her to adjust her analysis method to focus on the relationships between different attributes and her experimental data. In her next analysis phase, she investigated how temperature affects each alloy combination (Fig \ref{fig:case:large_projection_temp_color}), focusing on attributes that offer insights into the Pair Distribution Function (PDF) plots. These plots helped her link local atomic structures, seen as peaks in the PDF, to the properties of various alloys.
By examining the projection's top curve to see how values change within interesting loop shapes, she confirmed that with increasing temperature, the data points tend to converge (Fig. \ref{fig:case_study:merging_curve}-A).

She also makes note of the small, isolated worm-like clusters outside of the larger shape.
Drawing a shape over one of the small clusters, she can see that these small clusters are composed of a single temperature (Fig. \ref{fig:case_study:merging_curve}-B). She noted,
``well, something is definitely happening here! For the [\ce{Mn_{0.3}Ge_{0.7}Te}] datasets, it looks like it's making worms grouped by temperature for a bunch of different POSCARs. I assume that's the case for the [\ce{Mn_{0.25}Ge_{0.75}Te}], and [\ce{Mn_{0.2}Ge_{0.}Te}], worms as well.''
Color coding the projection by combination, she can confirm her assumption on the data subsets.

\begin{figure}[t!]
    \centering
    {\includegraphics[width=\linewidth]{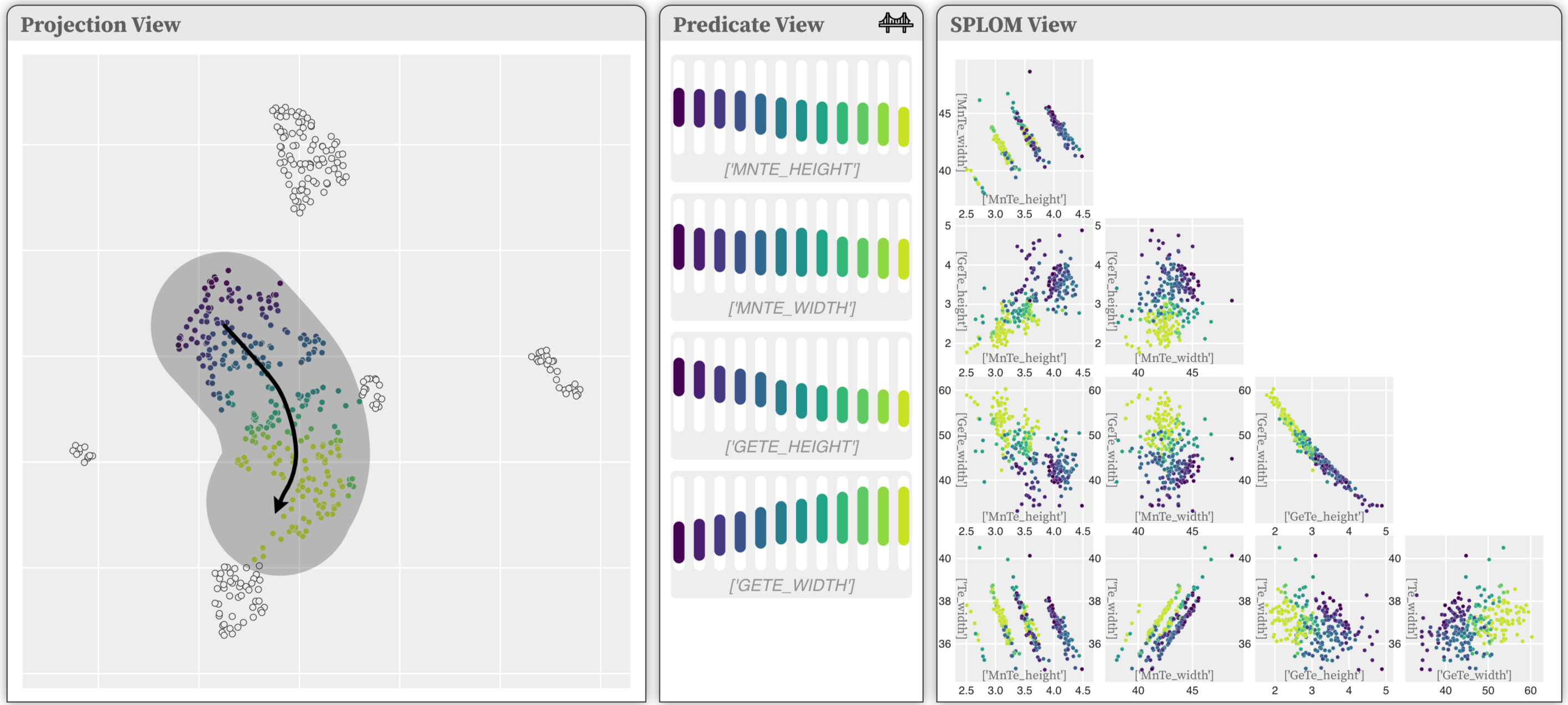}}
    \caption{\sys views showing the attributes of the drawn shape selection of the larger cluster. In the SPLOM, you can see the correlation between Te height and Mn height. }%
    \label{fig:case_study:MnTe_peak_influence_trail}%
\end{figure}

\subsubsection{Phase Three: Investigating Subsets of Interest}
In the next stage of her exploration, our collaborator creates a dataset for these subsets in question. Instead of the temperature dependencies, she focuses on the outputs of her existing model to get a clearer tie back to her experiments.
This dataset, smaller and more targeted, focuses on subsets that previously formed unique, worm-like clusters. Building on her analysis suggesting Mn influences Te positions, she first colors the projection by \ce{MnGe_r}. After identifying an interesting cluster, she examines its predicate ranges and potential relationships. Upon observing correlations, she draws a shape down the larger cluster (Fig. \ref{fig:case_study:MnTe_peak_influence_trail}), leading to a key observation: the height and shape of MnTe peaks significantly impact Te peak locations and heights, showing a stronger correlation with MnTe peaks than with GeTe/Te peaks.
This consistency across clusters suggests that while germanium (Ge) scarcely affects Te atom positions, manganese (Mn) plays a pivotal role, altering them noticeably.
``I really love seeing this because it's showing me how influential the Mn is in the PDF patterns.'' 

Revisiting the questions posed in the previous section:
\textbf{How does changing the proportion of manganese (Mn) affect the way the alloy's structure shifts from a slanted, diamond-like shape to a straight-edged, cube-like shape?} 
Our collaborator found that Mn significantly influences the alloy's structure, noting a clear correlation between the intensities of Mn-Te and Te-Te peaks across the alloy space. This observation aids in understanding how Ge atoms cause local distortions, affecting the alloy's appearance, especially in low Mn concentrations, where Ge tends to shift the structure towards a more rhombohedral form.
\textbf{Are the shapes formed by manganese (Mn) atoms similar across the alloy space?}
In the initial exploration stage, projections showed clustering by the number of Mn atoms in the coordination shell, regardless of composition. This suggests that the proximity of Mn atoms within their shells is more crucial than the overall elemental mix in the unit.

\subsection{Observations}
We observed a few notable things during our collaborator's interaction with \sys. Her process was highly iterative, moving from broad to more pointed in the data and hypotheses, often switching between several datasets. The adaptability of \sys proved to be a significant asset, for switching between one dataset to another, adding and subtracting columns used in the projection and data view from the system's UI, and getting immediate feedback on selections of interest. Particularly important was the ability to dynamically update the projection and data spaces, as this allowed her to rapidly iterate on her hypotheses and direction of analysis, as well as account for unexpected things, such as a cluster defined by attributes that were not expected. 
As well, our collaborator found \sys's functionality to draw a shape was extremely valuable for understanding the change in properties between one data subset to another, especially during the second phase of the analysis of temperature dependencies as seen in Figure \ref{fig:case_study:merging_curve}-A.

\section{Discussion}
\sys is designed to support the user in making sense of visual patterns in dimensionality reduction projections.
In this section, we reflect on emergent insights, implications of \sys's value beyond its original design goals, limitations, and some opportunities for future work.
\subsection{\sys Design Considerations: Why SPLOM?}



We chose to use a Scatter Plot Matrix (SPLOM) due to its efficacy for visualizing patterns in high dimensional space. This could involve the characterization of subspaces or the relationships between attribute values, which are both important tasks for our collaborators. Additionally, the SPLOM view allows you to visualize the predicate value ranges of the relevant attributes in all possible pairings. For these reasons, we chose SPLOM over alternative high dimensional techniques, detailed in \ref{subsec:related-highd}. However, future work will explore other methods of visualization to make \sys more flexible for a wider variety of datasets and tasks.

\subsection{The Value of Flexibility}
Along with creating a flexible analysis environment, it is also important to consider how we can enhance \sys to be more adaptable in terms of its composition and features. Below, we outline several opportunities for doing so.

\smallskip
\noindent \textbf{Beyond Projection Visualizations:} Not all data and tasks benefit from the exploration of a projection.
The projection view can be substituted for other methods of visualizing an overview or summary of the dataset, depending on the analysis tasks. 
Considering an example in the context of material science, imagine researchers are seeking to develop a new alloy with high strength and corrosion resistance for aerospace applications. They use a connectivity matrix to explore potential candidates, where the matrix includes a variety of known alloys, each characterized by their mechanical and chemical properties. Selecting cells or submatrices are entry points for understanding the properties of nearest neighbors of a selected cell, or defining ranges of desired properties to highlight relevant cells in the matrix.
Or in a more general context, one could also imagine a scenario where the projection view might contain geospatial data, and patterns of location could be explored across relevant dimensions.
Future work will explore the use of predicates as a bridge between high-dimensional data and alternative low-dimensional representations, beyond just projections.

\smallskip
\noindent \textbf{Beyond Axis-Aligned Dimensions:} \sys currently assumes the original (axis-aligned) data dimensions in the data-space visualization as they are the most interpretable.
However, this design requirement can be lifted for advanced users who can understand complex data dimensions.
Although we illustrated \sys using the original data dimensions in Section \ref{sec:casestudies}, we observe that the predicate induction could also have been performed using the principal components with the SPLOM showing the data with principal components as the data dimensions.
The resulting SPLOM visualization will be less interpretable, but the use of principle components (and possibly other non-linear dimensions) can result in predicates that better fit the user-selected data.

\smallskip
\noindent \textbf{Adapting Predicate Induction:}
\sys's predicate induction engine uses the RPI and \alg algorithms, representing two extremes of the accuracy/scalability tradeoff.
Future work exploring predicate induction algorithms that take a more balanced approach to this tradeoff could be beneficial.
Additionally, the predicate induction algorithm can be further tuned to the task of understanding DR results by explicitly accounting for distortions introduced by the DR algorithm.
``Distortion aware'' predicate induction could more effectively identify patterns in the original dimensions by adjusting for distortions in regions brushed by a user.

\section{Conclusion}
In this paper we present \sys, a system that \textit{bridge}s a projection space with the original data space using first-order predicate logic.
\sys connects patterns observed in the projection space to the original data space, helping users understand the pattern within the familiar data space. 
This decreases the likelihood of false discoveries resulting from spurious structure within the projection.
\sys is agnostic to the projection algorithm, the visualization technique used within the data space, and the predicate induction algorithm itself.
We illustrate three showcases of \sys within scientific data, motion-capture data, and imagery data.
Finally, we evaluated the utility of \sys with a domain expert who found the design to be helpful in her workflow.

\setlength\intextsep{0pt}



\bibliographystyle{abbrv}

 \bibliography{main, matt-refs}
\end{document}